\documentclass{aa}  

\usepackage{graphicx}
\usepackage{natbib}
\bibpunct{(}{)}{;}{a}{}{,}
\usepackage{txfonts}
\begin{document}

   \titlerunning{Sunspot positions, areas, and tilt angles from Scheiner, 1611--1631}
   \authorrunning{R. Arlt et al.}

   \title{Sunspot positions, areas, and group tilt angles for 1611--1631\\ from observations by Christoph Scheiner}

   \author{R. Arlt\inst{1}
          \and
          V. Senthamizh Pavai\inst{1,2}
          \and
          C. Schmiel\inst{1}
          \and
          F. Spada\inst{1}
          }

   \institute{Leibniz-Institut f\"ur Astrophysik Potsdam (AIP), 
		An der Sternwarte 16, D-14482 Potsdam\\
              \email{rarlt@aip.de} \and
             Institut f\"ur Physik und Astronomie, Universit\"at Potsdam, Karl-Liebknecht-Str. 24/25, 14476 Potsdam 
   }
   \date{Received 2014; accepted 2014}

 
  \abstract
{}
{ Digital images of the observations printed in the books
 ``Rosa Ursina sive solis'' and ``Prodromus pro sole mobili'' by Christoph Scheiner as well
  as the drawings from Scheiner's letters to Marcus Welser
  are analysed in order to obtain information on positions 
  and sizes of sunspots that appeared before the Maunder minimum.
}
{ In most cases, the given orientation of the ecliptic is 
  used to set up the heliographic coordinate system for 
  the drawings. Positions and sizes are measured manually 
  on the screen. Very early drawings have no indication
  of their orientation. A rotational matching using common
  spots of adjacent days is used in some cases, while in 
  other cases, the assumption of images being aligned with
  a zenith--horizon coordinate system appeared to be the
  most probable.
}
{In total, 8167 sunspots were measured. A distribution of 
 sunspot latitudes versus time (butterfly diagram) is 
 obtained for Scheiner's observations. The observations of
 1611 are very inaccurate, the drawings of 1612 have at least
 an indication of their orientation, while the remaining part 
 of the spot positions from 1618--1631 have good to very good
 accuracy. We also computed 697~tilt angles of apparently
 bipolar sunspot groups observed in the period 1618--1631.
 We find that the average tilt angle of nearly 4~degrees
 is not significantly different from 20th-century values.
}
{}

   \maketitle

%

\section{Introduction}

Solar activity is to a large extent characterized by the sunspot
number and the latitudinal distribution of spots as functions of
time. More information on the activity is of course accessible if 
enough details of the structures at the solar surface are available.

Extending the sunspot record back in time is not only a matter of
obtaining reliable sunspot numbers and related indices for as many
cycles as possible, but it is also a matter of reconstructing the
solar butterfly diagram of the Sun. With the available sources of
pre-photographic observations, we may be able to compile an almost 
complete butterfly diagram for the telescopic era since AD~1610.

There are not many publications with positional measurements of
sunspots observed in the beginning of this era, namely in the
first half of the 17th century. Studies of the solar rotation
are available for Harriot by \citet{herr1978} for the period
1611--1613, for Scheiner by \citet{eddy_ea1977} for the period
1625--1626, for Hevelius by \citet{eddy_ea1976} and 
\citet{abarbanell_woehl1981} for 1642--1644,
and for Scheiner as well as Hevelius by \citet{yallop_ea1982},
but the sunspot positions were not published.  Sunspot positions 
from Galileo's observations in 1612 were derived and made available
by \citet{casas_ea2006}\footnote{Historical archive of sunspot observations at http://haso.unex.es/}.

Christoph Scheiner lived from 1573--1650 
\citep{braunmuehl,hockey_ea2007}\footnote{Some biographies also 
give 1575 as the year of birth, e.g. \citet{brockhaus1992}} and 
was a member of the Jesuit society (Societas Iesu). The present 
paper is based on the copy of \cite{scheiner1630} stored in the
library of the Leibniz Institute for Astrophysics Potsdam (AIP)
and \cite{scheiner1651} at the library of ETH Z\"urich, 
available as a high-resolution digital version through the
Swiss platform for digitized content, 
e-rara\footnote{http://www.e-rara.ch/zut/wihibe/content/titleinfo/765922}.
The observations cover the years of 1611--1631, albeit most of the
data comes from 1625 and 1626. Data from the period before the
Maunder minimum are interesting as they may tell us
details about how the Sun went into a low-activity phase lasting
for about five decades, in particular since \citet{vaquero_trigo2015} suggested 
that the declining phase of the solar cycle started already near 1618.

If information on individual sunspots and pores within sunspot
groups is preserved, quantities such as the polarity
separation and the tilt angle of bipolar groups may be inferred.
It will be of particular interest whether these quantities 
behaved differently in the period before the Maunder minimum as
compared to the cycles after it or present cycles. This
is not the first attempt to use sunspot positions derived from
Scheiner's observations. An estimate of the differential rotation
was, for example, derived from Scheiner's book by \citet{eddy_ea1977}
who concluded that the rotation profile was not significantly
different from the one obtained for the 20th century. The authors 
did not publish the obtained spot positions though. Scheiner
actually noticed different rotation periods for different spots
spanning from 25~d to 28~d \citep[][p.~559]{scheiner1630}. He
does not, however, say that those periods were attached to
specific latitudes of spots.

Scheiner (1573--1650) belongs -- together with Johannes Fabricius, Galileo Galilei, 
Thomas Harriot, Joachim Jungius, Simon Marius, 
and Adam Tanner \citep{neuhaeuser2016} -- to the
first observers who left records of sunspots seen through a
telescope. Scheiner's first observations were made in early
March 1611, together with the student Johann Baptist Cysat (1586--1657). 
Scheiner was the first observer to start sunspot drawings on a
systematic, day-by-day basis in October 1611.
A comprehensive investigation of the sunspot observations of the
1610s can be found in \cite{neuhaeuser2016}.

The following Paper describes the drawings and gives details of the
drawings in Section~\ref{description}, explains the measuring
method for 1618--1631 in Section~\ref{coordinate_system}, the 
methods used to utilize the drawings of 1611--1612 in Section~\ref{1611},
evaluates the accuracy of the data in Section~\ref{accuracy}, 
gives with the data format and the butterfly diagram in Section~\ref{results}, 
and deals with the sunspot group tilt angles in Section~\ref{tiltangles}.
A summary is given in Section~\ref{summary}.


\section{Description of the Rosa Ursina and Prodromus drawings}\label{description}

\cite{scheiner1630} gives 71~main image plates with observations,
of which one is unnumbered as it contains only detailed faculae
drawings, plus an additional plate with his first observations of 1611.
The plates always contain information on more than one day. The
number of different days covered varies between four and 54. Early 
drawings are available from the letters of Scheiner to Marcus Welser
in 1611 and 1612. 
Further sunspot drawings are printed in \cite{scheiner1651},
where Scheiner defends the geocentric model. There are twelve image
plates with spots from 1625, 1626, 1629, and 1631. The drawings are in
exactly the same style as the ones in \citet{scheiner1630}, and
we used the same analysis procedure as described below.
The total 
number of days for which we have sunspot information is 798, while
two additional days show only faculae.
Table~\ref{annual} gives the annual numbers of days available.

\begin{table}
\caption{Numbers of days available for given years as reported
by \cite{scheiner1630}, \citet{scheiner1651}, and \citet{reeves2010}.\label{annual}}
\begin{tabular}{p{1.3cm}r|p{1.3cm}r|p{1.3cm}r}
\hline \hline
Year & Days & Year & Days & Year & Days\\
\hline
1611 & 41\tablefootmark{a}           & 1622 & 17\phantom{\tablefootmark{a}} & 1626 & 169\tablefootmark{b}\\
1612 & 31\phantom{\tablefootmark{a}} & 1623 &  9\phantom{\tablefootmark{a}} & 1627 & 55\phantom{\tablefootmark{a}}\\
1618 & 7\phantom{\tablefootmark{a}}  & 1624 & 40\phantom{\tablefootmark{a}} & 1629 & 49\tablefootmark{c}\\
1621 & 27\phantom{\tablefootmark{a}} & 1625 &343\tablefootmark{b}           & 1631 & 12\phantom{\tablefootmark{a}}\\
\hline
Total & \multicolumn{5}{l}{800}
\end{tabular}
\tablefoottext{a}{Five days were not analysed because of unreliable positional results.}
\tablefoottext{b}{One day showed only faculae.}
\tablefoottext{c}{Two days were omitted because groups were not drawn completely.}
\end{table}

Five observing locations contributed to this compilation of results 
on sunspots. Scheiner was located in Rome, Italy, in the years 
1624--1633 and we assume a geographical position of $12\fdg45$ eastern 
longitude and $41\fdg90$ northern latitude for his place. The other
locations were Ingolstadt in Germany, Douai (Duacum in Latin) in 
France, Freiburg im Breisgau, Germany, and Vienna, Austria. 

The University of Freiburg and the Jesuit church
next to it were located at $7\fdg85$~E, $48\fdg00$~N.   
One probable observer was a pupil of Scheiner, Georg Sch\"onberger
(or Schomberger) \citep{braunmuehl} who later observed with Johann Nikolaus Smogulecz (or
Jan Miko\l{}aj Smogulecki). The observations are also 
available in a separate publication by \citet{smogulecz_schoenberger1626}.

For Ingolstadt, we assumed the position of the original university building 
``Hohe Schule'' at $11\fdg42$~E and $48\fdg76$~N. 
Scheiner gave the geographical latitude on top of a page of 1611 
observations as $48\fdg67$. After Scheiner left Ingolstadt in 1616, the 
observers could have been Scheiner's pupils, among them Johann Baptist
Cysat, Chrysostomus Gall, and Georg Sch\"onberger \citep{braunmuehl}. Cysat taught
in Luzern, Switzerland, from 1624--1627 though \citep{bio}.
Since the observations from Ingolstadt that were not made by 
Scheiner cover the period from 1623 Mar~26 to 1625 Sep~15,
they were most likely not made by Cysat. Scheiner gives
Georg Sch\"onberger as the correspondent of the Ingolstadt
observations, but Sch\"onberger taught in Freiburg later, and his books of 1622 and 1626 were
published in Freiburg. It remains unclear to us who
was the observer in Ingolstadt.

For Douai, we assume approximately $3\fdg1$~E, $50\fdg4$~N.
Since the University of Douai was a conglomerate of colleges
and the Jesuits erected their own school in the 1620s, it is 
less obvious from which exact place they observed, but not 
really relevant for the scope of this Paper. The observer in 
Douai was Karel Malapert (Charles Malapert, Carolus Malapertius) 
who published his own observations in two books \citep{malapert1620,malapert1633}.
While it was published in Douai, \cite{braunmuehl} mentions his
observations to be made at Danzig, Poland. This may be a
translation error of Duacum, although Malapert did teach in Poland, but
at the Jesuit College of Kalisz from 1613--1617 \citep{birkenmajer1967}. 
Scheiner regularly compares the observations of the
three sites in his figures. We include all spots visible in
these drawings in the data set, but if a spot was observed 
by Scheiner as well as a colleague on the same day, we give 
only Scheiner's spot position.

The location of the Jesuit church in Vienna is about
$16\fdg4$~E, $48\fdg2$~N. Scheiner names Johann Cysat as
the observer whose life between 1627 and 1631 is not
known in detail. At least, we find an indication that Cysat was
in Vienna ``occasionally'' in that period from \citet{bio}.
Scheiner compared the Vienna data with the Rome data, so there
is eventually only a total of six spot measurements that complemented
the Rome data (recorded on 1629 Aug~12, 13, and 20).

Typical figures contain a circle for the solar limb, a horizontal
line mostly denoting the ecliptic and a selected number of
sunspot groups which are followed on several days. The dates
(in many cases together with a precise time and the elevation
of the Sun) are given in a small table within the circle.

The next figure gives another set of selected groups for a sequence
of days. Note that these sequences of some drawings very often overlap.
The figures are made in a way as to show all appearances
of a given group on as many days as necessary.

Times are given in 12-hour format, annotated with ``m'' = matutinus 
= morning and ``u'' = uespera = vespera = evening. The only exception
is the first drawing of the 1624--1631 spell. The times are all above 
12:00, but since Scheiner also gave the elevations of the Sun, we can 
guess a most plausible meaning for the times. We find that the time was 
very likely measured from the last sunset, which is at about 16:25 local
time in Rome in mid-December. The corresponding local times in
today's reading are given in Table~\ref{plate1}, sorted by local
time instead of date. All other plates give local times apparently
measured from midnight.

\begin{table}
\caption{Assumed times specially for Plate~I.\label{plate1}}
\centering\begin{tabular}{lcc}
\hline \hline
Printed time & Assumed time & Printed elevation \\
\hline
Dec 17, a.m. 16:48        &  09:13       & $14\fdg5$ \\
Dec 19, a.m. 17:30        &  09:55       & $17\fdg7$ \\
Dec 16, a.m. 18:48        &  11:13       & $24\fdg5$ \\
Dec 16, p.m. 21:00        &  13:25       & $21\fdg7$ \\
Dec 14, p.m. 21:45        &  14:10       & $22\fdg5$ \\
\hline
\end{tabular}
\end{table}

The printed versions of the drawings were digitized with a book
scanner at a resolution of 200~dpi. One pixel corresponds to
0.13~mm converting to an angular distance of a bit less
than $0\fdg1$ in heliographic coordinates in the solar disk centre.
The observations of Dec~14, 1611 to Apr~7, 1612 were
not shown by \citet{scheiner1630}; these were digitized from
\citet{reeves2010}. Their diameter is about 22.7~mm. A
distance of 0.1~mm on those disks corresponds to an angle of
$0\fdg5$ in heliographic coordinates in the disk centre.

\begin{figure}
\centering
\includegraphics[width=0.485\textwidth]{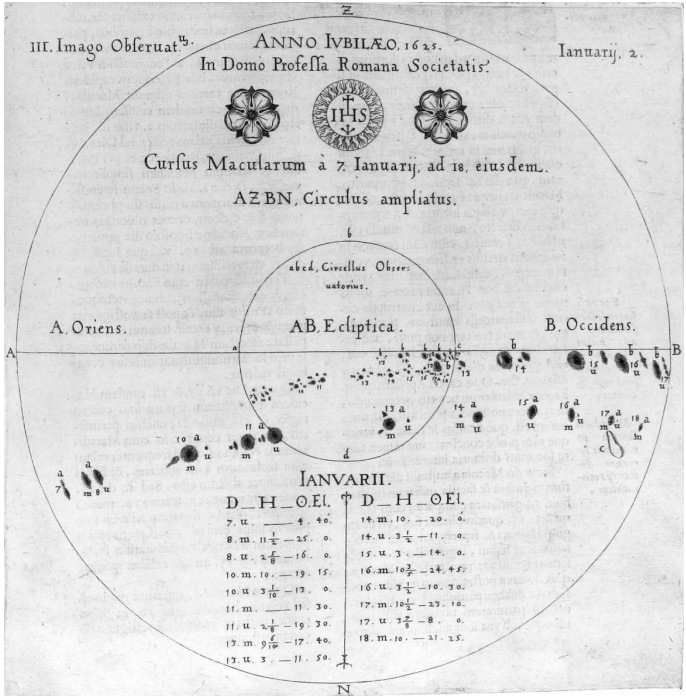}
\caption{Example of an early drawing indicating the directions and an inner circle
annotated as the observed circle. All spots appear twice: once in the
inner circle and once in the full circle.
\label{scheiner_024}}
\end{figure}


\begin{figure}
\centering
\includegraphics[width=0.485\textwidth]{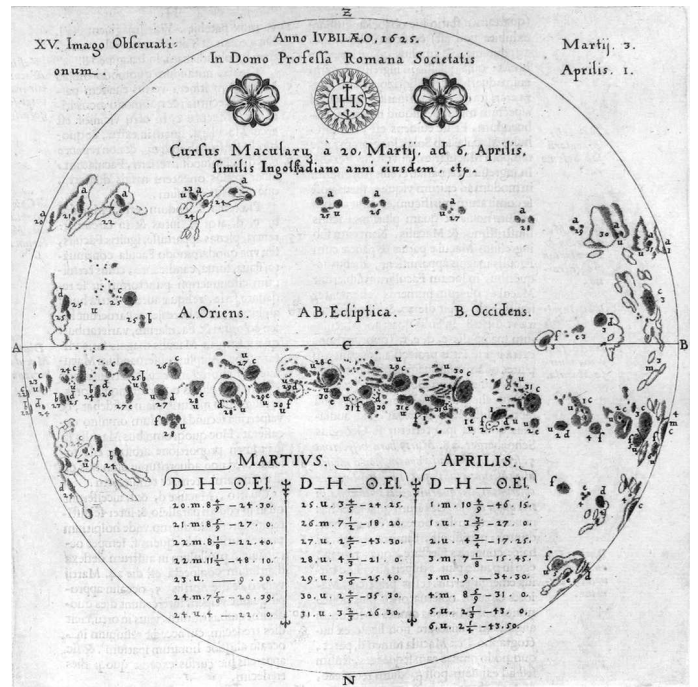}
\caption{Example of a variable ecliptic line to avoid crowdedness. Spots
near the eastern and western limbs of the southern-hemisphere groups are
rotated above and below their actual position for better visibility.
Group ``a'' in the northern hemisphere is unaffected.
\label{scheiner_035}}
\end{figure}

The images~I to~V, IX, X, XVI, XVII contain a smaller circle which is 
annotated as the ``circellus observatorius'' = ``observed circle'', 
while the large circle covering almost the entire page is entitled the 
``circulus ampliatus ex observatorio'' = ``circle expanded by observatory.''
Fig.~\ref{scheiner_024} shows the example of Plate~III. Spots are
plotted twice in such drawings, once at the scale of the inner
circle and once at the scale of the outer circle. We use the large
version of the drawing for position and area measurements. As shown
in Sect.~\ref{accuracy}, the accuracy of the positions is remarkable,
even though the large images may be the result of post-observational
magnifications.

Another peculiarity is shown in Fig.~\ref{scheiner_035} covering the
period of 1625 March~20 to April~6. The southern hemisphere of the 
Sun was so crowded with spots that some sunspot groups were plotted
into a rotated coordinate system to avoid overlap with groups of
other days. A few spots were actually plotted twice: once within the
crowd of spots and once at another location. These spots indicate
that it is indeed a rotation of the disk by multiples of roughly
$15\degr$ that led to the secondary positions, and not a linear
shift upward or downward.

Image~XX contains a large number of faculae crossing the solar
disk, while there is only a small group on the southern hemisphere
for which we actually determined positions and sizes.

Image~XXII shows extended faculae at the eastern limb and sunspots
for the period 1625 May~2--14. The faculae of May~3 are drawn in
rather dark colour, possibly indicating very bright faculae,
keeping in mind that faculae needed to be plotted with an inverse
grey scale. There is a large c-shaped black area surrounding a
blank-paper region which appears ``bright'' in contrast to the
``dark'' surrounding faculae. \citet{rek2010} seem to argue in a footnote
that this region may have been a white-light flare. The
textual description of May~2--4, 1625, by \citet[][p.~208]{scheiner1630} 
says the following: 
\begin{quote}
 On the second day of March, two spots appeared which were
 preceded by a facula. On the third day, they showed up with a much
 more luminous retinue of faculae and a more pompous armament of
 shadows blending into spots. Just as the faculae extended on
 the forth day and the shadows dissipated, many more spots appeared.
\end{quote}
(While Scheiner uses the word ``umbra'', we translated it into shadow, 
since the conception of an umbra may have been different from 
today's definition.) Since there is no mention of a phenomenon being 
variable during the course of the day, there is no strong evidence for a white-light flare 
despite the presence of a peculiar sunspot group.

There is an unnumbered image plate between images~XXVI and~XXVII
which contains only faculae and covers days that were already 
shown in images~XXIII and~XXIV and has therefore not been used in 
our measurements.


\section{The coordinate system of the drawings of 1618--1631}\label{coordinate_system}

Scheiner describes his observing method as a projection behind
the telescope(s). This would mean that the appearance of
the Sun was mirrored. There are several indications, 
however, that he made the actual images in a non-mirrored
way, and the orientation of the drawings is roughly upright.

More precisely, the drawings always show a nearly horizontal line 
representing the ecliptic. Since the plane of the ecliptic is not 
easily accessible when observing under the sky, it has apparently 
been computed by Scheiner from the direction to the local zenith. 
The direction to the local zenith is marked on a few drawings 
by little dots at the solar limb.

We need to find the angle between the direction of the solar
rotation axis and the pole of the ecliptic. As already done in 
\cite{arlt_ea2013}, we use the output of the angle of the solar 
rotation axis with the direction to the true-of-date celestial 
north pole as provided by the JPL Horizons ephemeris 
webpage\footnote{http://ssd.jpl.nasa.gov/horizons.cgi}. The ecliptic 
pole is assumed to be at $\alpha_{\rm E}=18^{\rm h}$, $\delta_{\rm E} = 66\fdg56$
which is true-of-date by definition (neglecting orbital precession and nutation).
We then use a spherical transformation to obtain the missing angle
between the celestial north pole and the ecliptic pole for a
given date and location. When transforming the celestial
coordinates of the ecliptic pole into a system with its pole in
the center of the Sun, the new longitudinal angle (new ``right
ascension'') is the desired angle. Let $\alpha_\odot$, $\delta_\odot$
be the celestial coordinates of the Sun and $\alpha_{\rm E}$
and $\delta_{\rm E}$ the celestial coordinates of the ecliptic
pole, then
\begin{eqnarray}
\sin\delta' &=&\cos\delta_{\rm E}\cos\alpha_{\rm E}\cos\delta_\odot + \sin\delta_{\rm E}\sin\delta_\odot\nonumber\\
\cos\hat\alpha &=& (\cos\delta_{\rm E}\cos\alpha_{\rm E}\sin\delta_\odot - \sin\delta_{\rm E}\cos\delta_\odot) / \cos\delta'\nonumber\\
y &=& \sin\alpha_{\rm E}\cos\delta_{\rm E}\nonumber\\
\alpha' &=& \left\{\begin{array}{ll}
  \pi-\hat\alpha\quad\mbox{~if~}y \geq 0\\
  \hat\alpha-\pi\quad\mbox{otherwise}\\
  \end{array}\right.
\\
\end{eqnarray}
Looking at the Sun, $\alpha'$ now is also an angle running counter-clockwise.

We can now cross-check the accuracy of Scheiner's determinations
of the ecliptic by computing the angle between the zenith and the
pole of the ecliptic. This angle can be measured on a few drawings
where the zenith is indicated at the solar limb. We are again using
the spherical transformation and replace the celestial coordinates
of the ecliptic pole by the celestial coordinates of the local
zenith at the time of observation. This position is provided by
the {\tt zenpos} routine of the IDL Astronomy User's Library 
of November 2006 \citep{landsman1993}. Table~\ref{angles} gives
a comparison of angles between the zenith and the north pole of
the ecliptic, one value being measured on the drawing directly,
and the other values being computed from the solar ephemeris.
The differences are typically a fraction of a degree, but reach
$1\fdg9$ on 1625 Nov 14.

\begin{figure}
\centering
\includegraphics[width=0.485\textwidth]{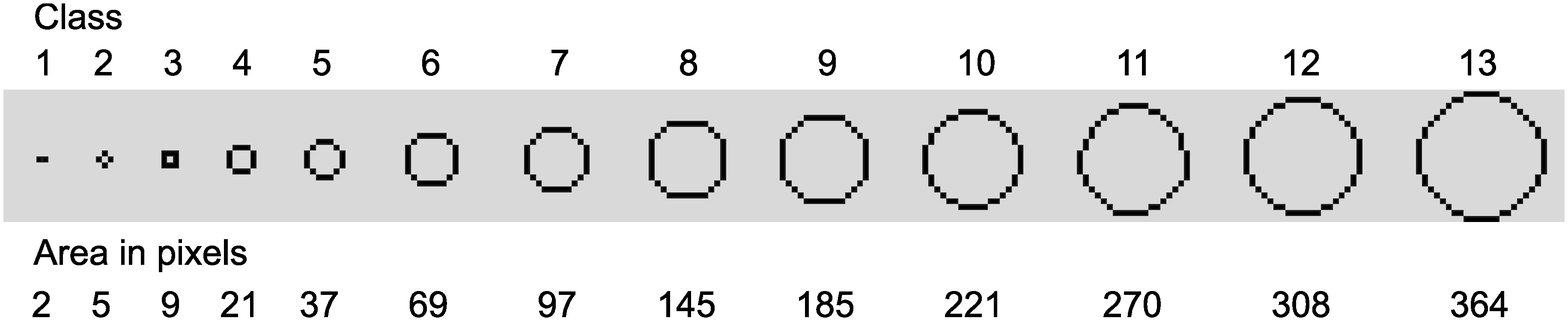}
\caption{Cursor shapes used to estimate the sizes of the spot umbrae
and pores.\label{cursor_shapes}}
\end{figure}

\begin{table}
\caption{Angle between the direction to the pole of the ecliptic and the zenith
as indicated in the drawings and theoretically determined. The last column is
the absolute difference between drawing and ephemeris.}
\label{angles}
\centering
\begin{tabular}{lrrr}
\hline \hline
Date and time       & Drawing & Ephemeris & Diff.\\
\hline
1625 May 19, $08^{\rm h}00^{\rm m}$   & $-38.5$   & $-38.9$ & $0\fdg4$\\
1625 May 20, $16^{\rm h}10^{\rm m}$   & $ 64.2$   & $ 65.0$ & $0\fdg8$\\
1625 May 28, $08^{\rm h}35^{\rm m}$   & $-41.2$   & $-42.1$ & $0\fdg9$\\
1625 May 31, $08^{\rm h}10^{\rm m}$   & $-43.2$   & $-44.4$ & $1\fdg2$\\
1625 Nov 03, $08^{\rm h}22^{\rm m}$   & $-54.7$   & $-55.1$ & $0\fdg4$\\
1625 Nov 12, $09^{\rm h}00^{\rm m}$   & $-47.6$   & $-47.0$ & $0\fdg6$\\
1625 Nov 14, $14^{\rm h}30^{\rm m}$   & $ 14.3$   & $ 16.2$ & $1\fdg9$\\
\hline
\end{tabular}
\end{table}

\begin{figure*}
\centering
\includegraphics[width=0.30\textwidth]{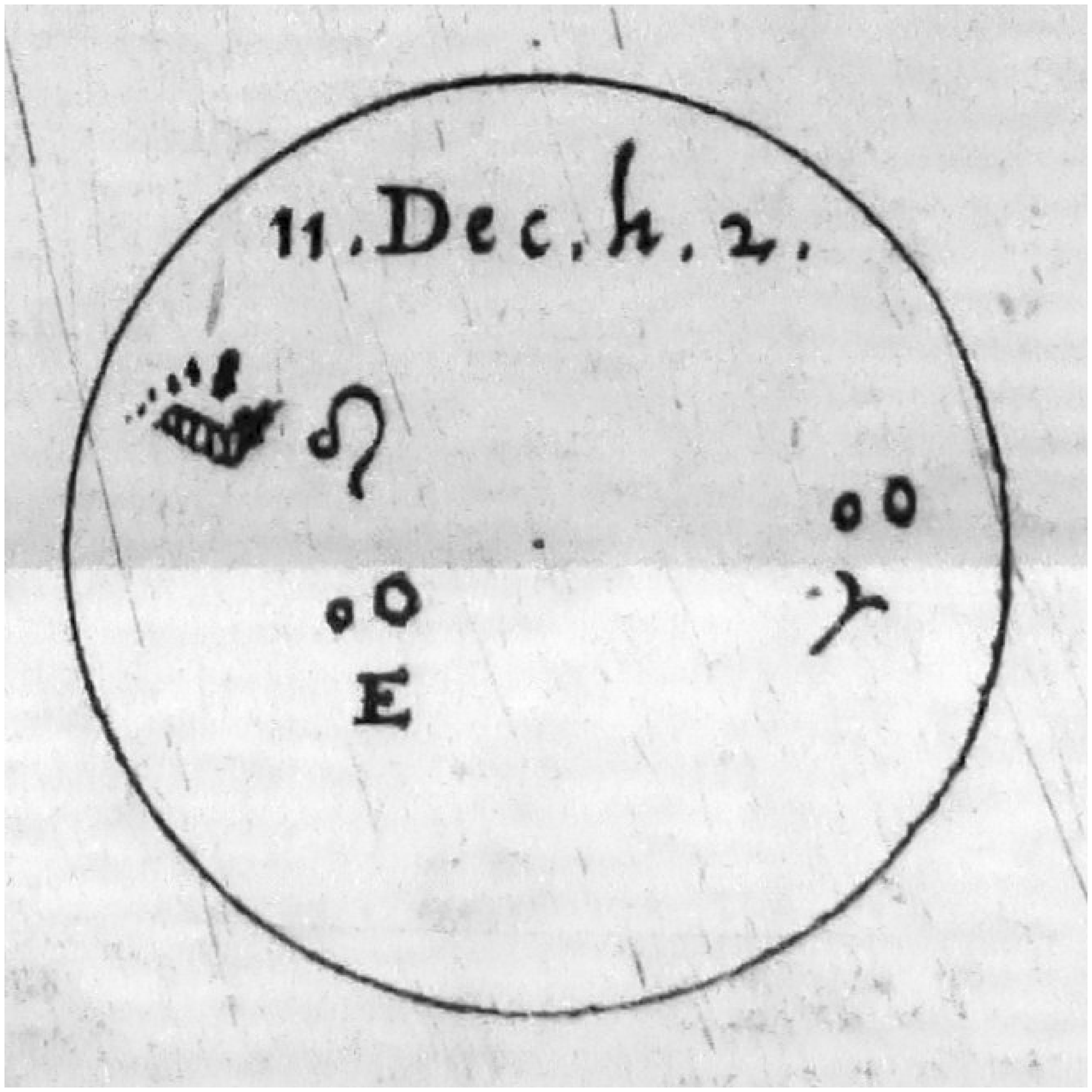}
\includegraphics[width=0.30\textwidth]{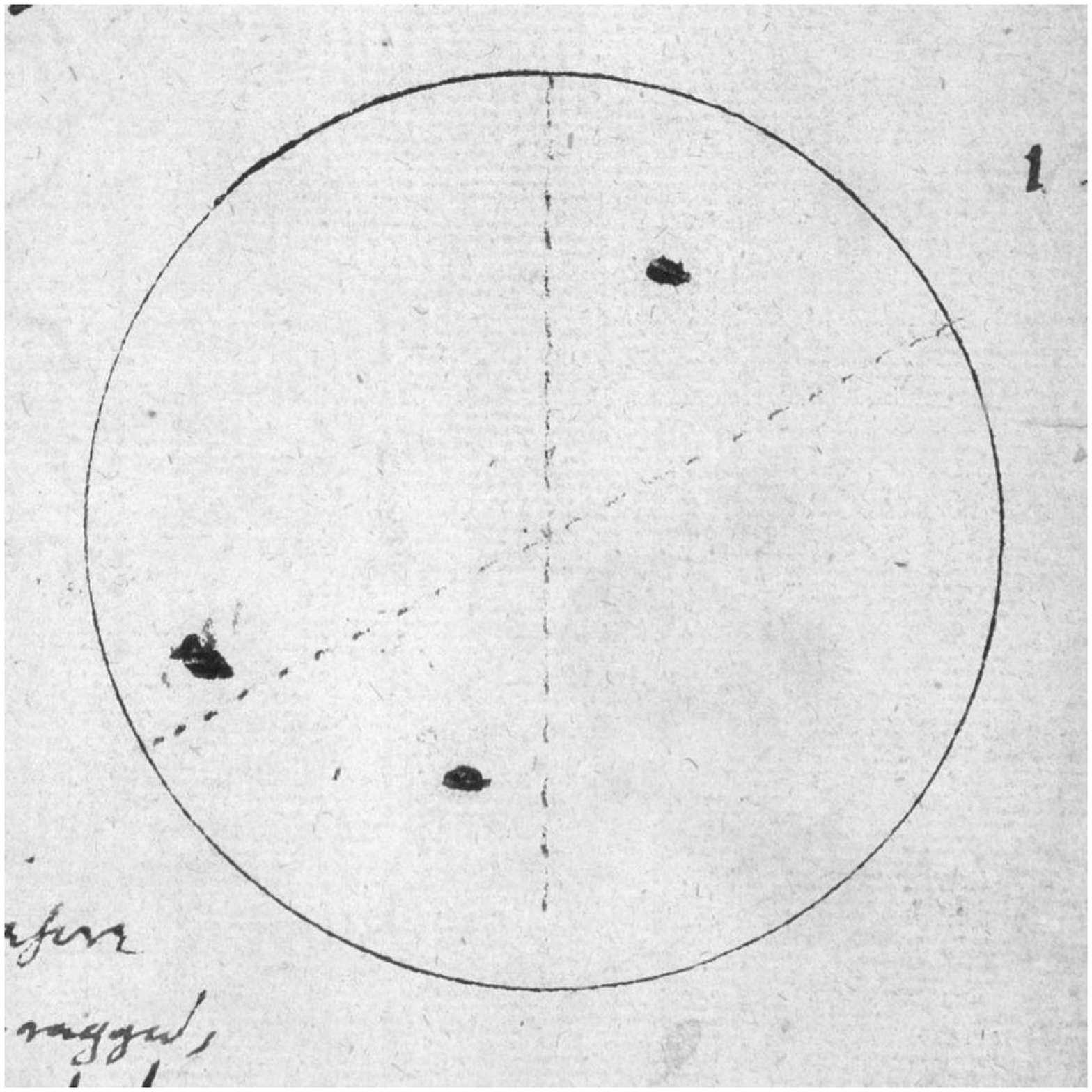}\\
\centering
\includegraphics[width=0.30\textwidth]{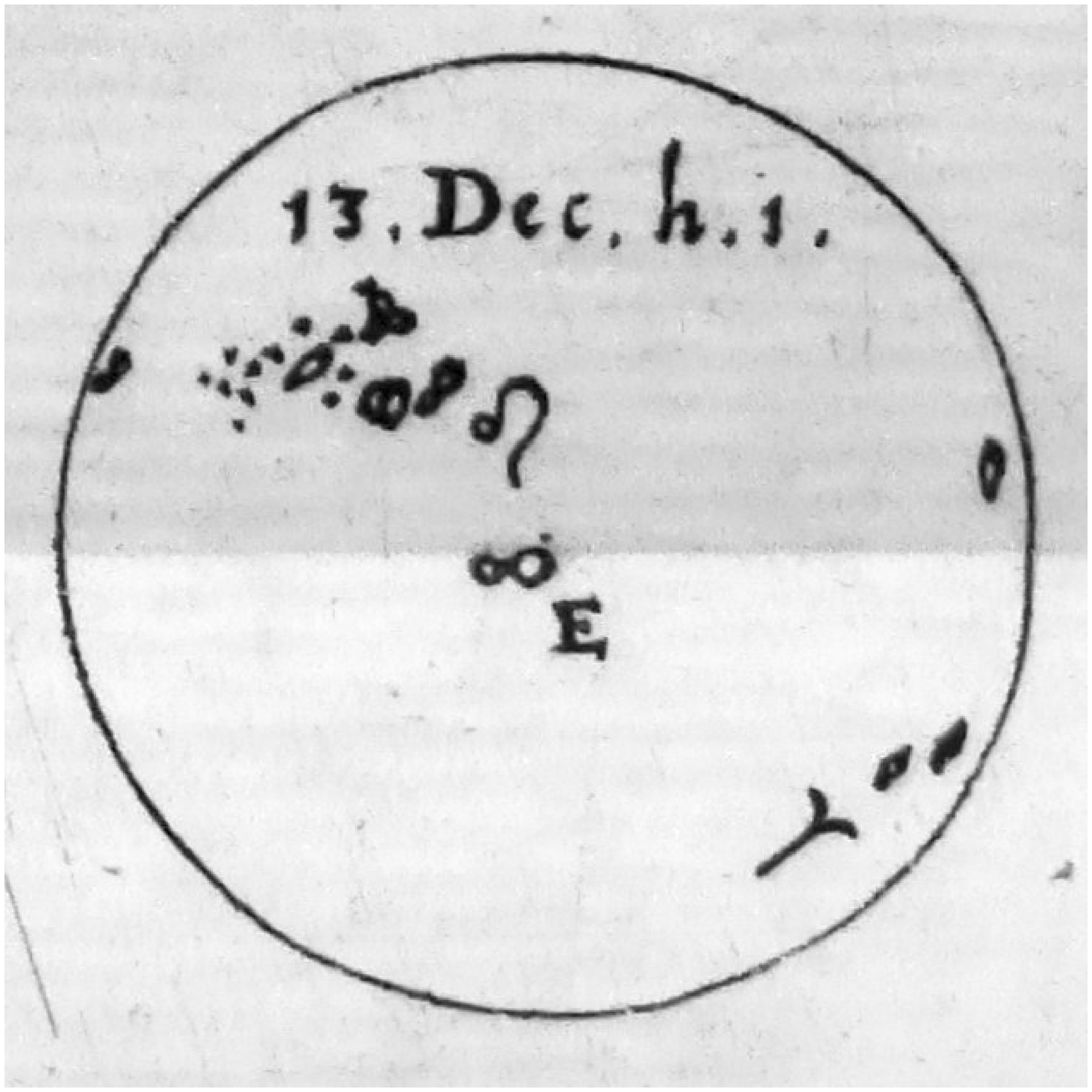}
\includegraphics[width=0.30\textwidth]{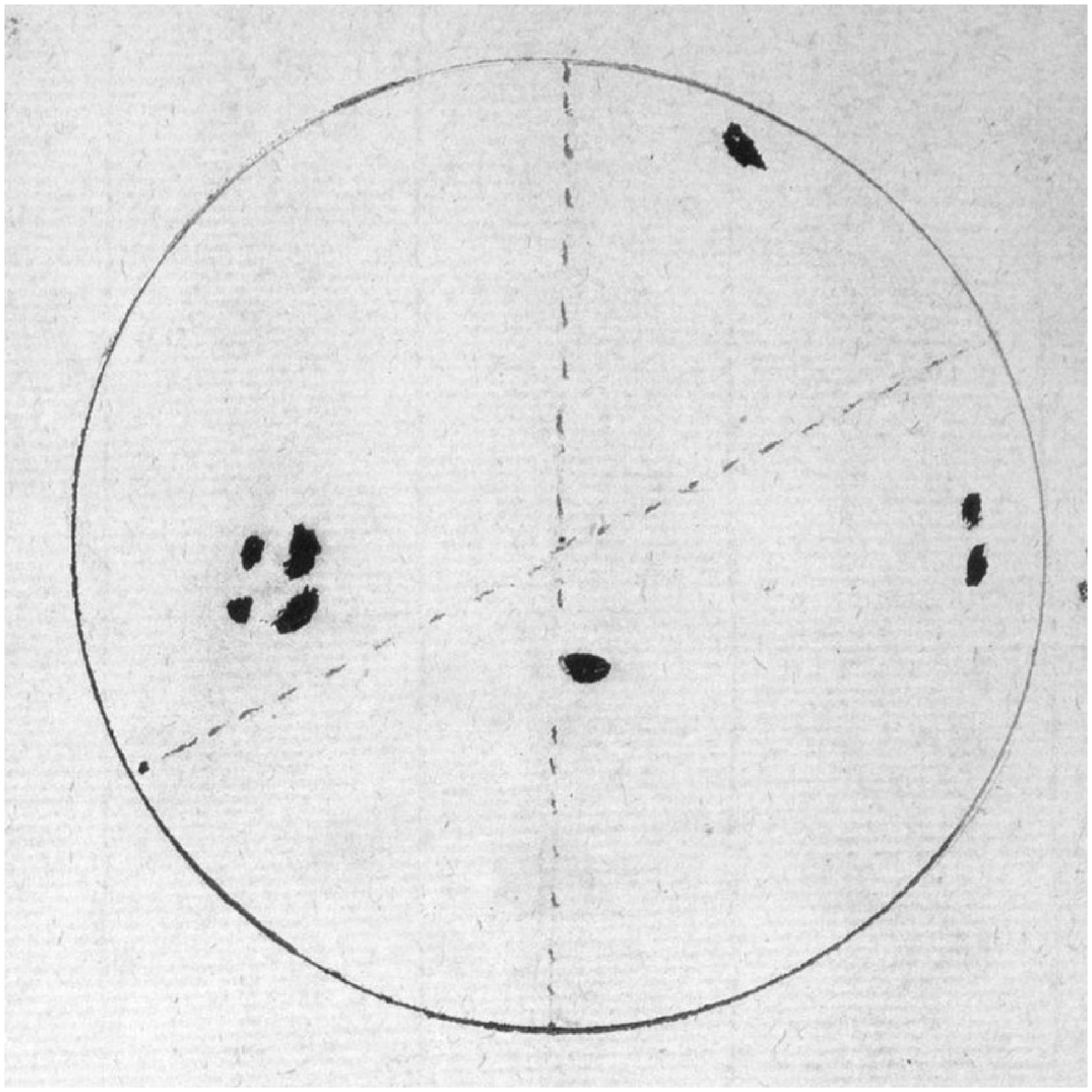}
\caption{Comparison of the drawings by Scheiner (left) and Harriot (right) on the
two days they have common observations. The top row shows the observations of
1611 December~11, at 14~h (left) and at 10~h (right). The bottom row is of
1611 December~13, at 13~h (left) and at 8:30~h (right). The vertical dashed 
lines in Harriot's drawings denote the (observed) direction to the zenith, the
other dashed lines are the (computed) ecliptic. \label{scheiner_harriot}}
\end{figure*}

Since neither the printing of the drawings nor the digitization process
ensure that the ecliptic is an exactly horizontal line in the
image, we also need to add the angle of the printed line which 
is obtained by clicking on two points on the line in the image
and computing the angle (typically below $1\degr$).

The actual process of measuring the sunspots consists of the following 
steps:
(i) cutting out the solar disk from the full image by
    clicking on the left-most, right-most, lowest and
    uppermost limb of the Sun; this also allows for a
    certain degree of ellipticity of the solar disk to
    be measured correctly),
(ii) determining the exact angle of the ecliptic
    line by two clicks, left and right,
(iii) setting up the spherical coordinate system as
    supported by IDL,
(iv) clicking on the relevant spots with thirteen different
    cursor sizes (Fig.~\ref{cursor_shapes}), for which the best 
    fit in position and size to the umbrae and pores are sought visually.

In the cases where a rotated coordinate system was used to draw
spots of crowded regions (see Sect.~\ref{description}), we can 
determine the  positions if at least one spot was plotted twice,
once in the standard ecliptical system, and once in the rotated
one. The angle with the disk center gives us the rotation of the
coordinate system for the displaced spots. For 15~spots, such 
a duplicate spot was not available, and we omit the positions
of these spots, but do keep corresponding records in the data
file.

Physical areas are then derived from the pixel area of the cursor
shapes $A_{\rm cur}$ used for the individual spots as compared with the total
pixel area of the solar disk, $A_{\rm disk}$. The heliocentric distance of the
spot from the disk center $\delta$ yields the correction for the
geometrical foreshortening, and we express the final area in millionths
of the solar hemisphere (MSH), i.e.
\begin{equation}
  A = \frac{1}{2\cos\delta}\frac{A_{\rm cur}}{A_{\rm disk}}\times 10^6.
\end{equation}

\section{The observations of 1611--1612\label{1611}}
The observations of 1611 October~21 to 1612 April~7 are 
Scheiner's first drawings of sunspot observations and are more difficult
to measure. The ones of 1611 October~21--December~14 are compiled
on a single page in the Rosa Ursina. The page is essentially the same as the one Scheiner
sent to the scientific friend Marcus Welser using the pseudonym 
{\em Apelles latens post tabulam\/} (Apelles hiding behind the scaffold) 
written on December~26, 1611, and published by Welser in January 1612, 
together with two earlier letters \citep{braunmuehl}.
Additional drawings until 1612 April~7 are available from the letters to Welser of 
1612 Jan~16, Apr~14, and Jul~25 \citep{scheiner1612}. We did not have access to the
originals, but used the reproductions of the drawings in \cite{reeves2010}
instead.

Some observations show an approximate orientation: the ones of the 
morning and afternoon of October~22 show the horizon, the ones from
1611 December~10 to 1612 January~11 show the ecliptic;
the other drawings have no information about the orientation of the 
solar disk. The ecliptic was drawn in connection with the
expected Venus transit on December~11 which was actually an upper
conjuction of Venus and the Sun. The images of December~10--13 also 
contain the expected path of Venus. The path is confusing as to whether
the images may be upside-down or mirrored, since it lies north of the
ecliptic while the true path was south of it. Since the spots are
clearly moving from left to right, there was still the possibility 
that the images are mirrored vertically (a view on a projection screen
behind a Galilean telescope). The description reveals, however, that
the path was indeed meant to be north of the ecliptic; the accuracy
of the ephemeris of Venus was just not good enough (especially of the ecliptic
latitude) to place it right at the time \citep{reeves2010}. For all
drawings showing the ecliptic, we computed the angle between the 
solar axis and the direction to the pole of the ecliptic and set up
the coordinate system in the same way as for the observations of 1618--1631.
There were 29~observations for which the ecliptic was used in 1611--1612.

There are three options for the analysis of the remainder of the drawings:
(i) using two or more drawings to fix their orientation with
the spot displacements due to the solar rotation (``rotational matching''); 
(ii) assuming that all the drawings are plotted in a horizontal
system, i.e. the vertical on the book page points to the zenith;
or (iii) choosing an arbitrary orientation such that the distribution 
of spots agrees with our today's expectation of sunspot latitudes.

The method (i) for the rotational matching is the same as described
in \cite{arlt_ea2013}. Two or more spots which can be identified
in two or more consecutive drawings are used with their 
measured Cartesian coordinates for a Bayesian inference of
their heliographic central-meridian distances and latitudes and
the orientation angles of the drawings. A differential rotation
as derived by \cite{balthasar_ea1986} is used for the solar
rotation with a fixed dependence $\Omega = (14.551 - 2.87\sin^2\lambda)$~$\degr$/day,
where $\lambda$ is the heliographic latitude.
The relation was computed from the sunspot group series of 1874--1976
compiled by the Royal Greenwich Observatory.

A Markov chain Monte Carlo search in the parameter space provides
us with full posterior probability density distributions for each
of the free parameters. In contrast to a best-fit search, they allow 
us to judge the quality of the rotational matching by the width
of the probability density distribution, its
skewness, or possible non-uniqueness of probable solutions.
The orientations of a total of 23~observations was fixed with
the rotational matching.

Since imposing the differential rotation of the Sun
to the solution of the orientation does not allow a subsequent
determination of the differential rotation in a possible future study,
we try to use method (ii) for as many cases as possible.
The celestial position of the Sun is provided by 
the Horizons ephemeris in a J2000.0 system. Since the angle between 
the solar axis and the direction to the celestial north pole
is given in a true-of-date system, we transform the Sun's
coordinates to a system of 1612 using the precess.pro routine 
from the IDL Astronomy User's Library. The position of the zenith is
computed by the zenpos.pro routine in a true-of-date system as well. 

The observations of March and April~1612 are the only ones with a reliable
alignment of the zenith direction with the vertical of the images.
This was shown by a comparison of measurements with the zenith-assumption
and measurements using the rotational matching. Table~\ref{comparison}
shows the average deviations of the spot positions between the
two methods. While the zenith-assumption delivered fairly consistent
results of spot motion across the solar disk, the rotational
matching yielded rather broad probability density distributions for the
position angles and we kept its results only for 1612 Mar~17, 19,
and Apr~1. In total, 27~observations were treated with this assumption of
a horizontal coordinate system.

Method (iii) was applied only when the first two methods led
to improbable spot distributions. We essentially applied a
position angle which minimizes the absolute latitudes of the
spots. Only four observations were treated with arbitrarily
chosen orientations (1611 Nov~7, 13, 14, and Dec~8).

The observations of 1611 Nov~6, 9, 10, 12, and Dec~24 were 
omitted entirely because no reasonable match with adjacent 
observations was possible and the spot distribution appears
to be highly improbable.

The resulting heliographic latitudes of the spots measured in
the 83~observations used from 67~days in 1611--1612 are between $-43\degr$ 
and $+42\degr$.

The sizes of the spots were strongly enlarged, and groups of smaller
spots were combined into one large spot, as described by Scheiner
(see \cite{reeves2010} for a translation). A size estimate can
only be arbitrary at this point, and we chose -- in order to be
able to use the 13~cursor shapes -- to select the size class
which matches roughly half the diameter of the plotted spot.
This choice still gives too large areas when converting the
disk area fraction of the cursor shape directly to MSH. The
areas in the data file are marked with ``!'' to indicate that
these need to be used with care (see Table~\ref{format}).

\begin{table}
\caption{Comparison of different coordinate systems adopted for
a selected set of observations in 1612.\label{comparison}}
\begin{tabular}{lrr}
\hline \hline
Date        &  Number & Avg. distance \\
            & of spots&               \\
\hline
1612 Mar 16 &    5    &  $0\fdg58$    \\
1612 Mar 17 &    5    &  $5\fdg80$    \\
1612 Mar 18 &    4    &  $0\fdg30$    \\
1612 Mar 19 &    5    &  $3\fdg63$    \\
1612 Apr 02 &   11    &  $7\fdg34$    \\
\hline
\end{tabular}
\end{table}

\section{Accuracy of the positions and areas\label{accuracy}}
In the first period of 1611--1612 when Scheiner drew the
sunspots with poor quality, there are actually two occasions
when Scheiner and Thomas \citet{harriot1613} observed on the same day (Fig.~\ref{scheiner_harriot}).
The images are rotated against each other, since Harriot
observed in the morning and Scheiner in the afternoon.
There is a fair agreement on the distribution of spots,
except that Scheiner noticed a spot near the eastern solar 
limb on Dec~13, 1611, which Harriot could not yet detect. Scheiner's
drawings are more detailed, whereas the spots are more
exaggerated in size than in Harriot's drawings. While the
drawings look qualitatively similar, we find positional
differences of up to $20\degr$ on those two days (assuming
Harriot's vertical lines are the direction to the local zenith).
This underlines the limited use of the spots in 1611.

Accidentally, Scheiner plotted the positions of one spot 
in two different image plates (XLVI, spot labelled as ``k'', 
and XLVII, spot labelled as ``a'') from 1625 
Nov~6--11. The differences give us an information about 
how well the positions could be reproduced in the plates
and are listed in Table~\ref{tab_accuracy}. The average
distance between the the positions of the spot in plate~XLVI 
and the corresponding positions in plate~XLVII is $0\fdg8$.
The average ``error'' of the cursor sizes chosen for the
various instances of the spot is 0.5 classes.
We kept the positions of plate~XLVII, since the spot continued to
exist for more days after Nov~11.

\begin{table}
\caption{Comparison of the positions of the same
spot in two different image plates.\label{tab_accuracy}}
\begin{tabular}{l|rrc|rrc|l}
\hline \hline
Date   &\multicolumn{3}{c}{Plate XLVI} &  \multicolumn{3}{c}{Plate XLVII} & \\
1625   & CMD   &  $\lambda$  &$S$ &CMD   & $\lambda$  &$S$ &$\delta$\\
\hline
Nov 06 &$-73\fdg9$& $-6\fdg0$&3  &$-74\fdg4$ &$-5\fdg8$ &3 & 0\fdg54\\
Nov 07 &$-60\fdg6$& $-5\fdg5$&5  &$-61\fdg6$ &$-6\fdg2$ &5 & 1\fdg22\\
Nov 08 &$-47\fdg1$& $-5\fdg9$&4  &$-48\fdg3$ &$-5\fdg9$ &5 & 1\fdg19\\
Nov 09 &$-33\fdg9$& $-5\fdg0$&5  &$-34\fdg8$ &$-5\fdg0$ &5 & 0\fdg89\\
Nov 10 &$-20\fdg4$& $-4\fdg7$&5  &$-20\fdg9$ &$-4\fdg8$ &4 & 0\fdg51\\
Nov 11 &$ -7\fdg4$ &$-4\fdg8$&5  &$ -7\fdg8$ &$-5\fdg0$ &4 & 0\fdg45\\
\hline
\end{tabular}
\tablefoot{CMD is the central-meridian distance, $\lambda$ is the
heliographic latitude, $S$ is the cursor size class according to 
Fig.~\ref{cursor_shapes}, and $\delta$ contains the heliocentric 
distance between the two measurements.}
\end{table}

\begin{table*}
\caption{Data format for the positions and areas of individual sunspots observed by
Christoph Scheiner and his collaborators.\label{format}}
\begin{tabular}{lllp{12.9cm}}
\hline \hline
Field & Column & Format & Explanation \\
\hline
{\tt YYYY}  & 1--4   & I4     & Year \\
{\tt MM}    & 6--7   & I2     & Month \\
{\tt DD}    & 9--10  & I2     & Day referring to the civil calendar running from midnight to midnight, Gregorian calendar\\
{\tt HH}    & 12--13 & I2     & Hour, times are mean local time at the observer's location\\
{\tt MI}    & 15--16 & I2     & Minute, typically accurate to 15~minutes\\
{\tt T}     & 18     & I1   & Indicates how accurate the time is. $T=0$ means the time has
                 been inferred by the measurer (in most cases to be 12h~local time);
                 $T=1$ means the time is as given by the observer; $T=2$ means the time was not printed, but inferred from the elevation of the Sun and the morning/afternoon discrimination given by the observer.\\
{\tt L0}    & 20--24 & F5.1   & Heliographic longitude of apparent disk centre seen from Rome\\
{\tt B0}    & 26--30 & F5.1   & Heliographic latitude of apparent disk centre seen from Rome\\
{\tt CMD}   & 32--36 & F5.1   & Central meridian distance, difference in longitude from disk centre;
                 contains NaN if position of spot could not be measured.\\
{\tt LLL.L} & 38--42 &F5.1 & Heliographic longitude in the Carrington rotation frame;
                 contains NaN if position of spot could not be measured.\\
{\tt BBB.B} & 44--48 & F5.1 & Heliographic latitude, southern latitudes are negative;
                 contains NaN if position of spot could not be measured.\\
{\tt M}     & 50    & C1   & Method of determining the orientation. `E': ecliptic present in drawing;
                 `H': book aligned with azimuth--elevation; `A': arbitrarily chosen orientation according to the 
                 distribution of groups; `Q': rotational
                 matching with other drawings (spots used for the matching have ${\rm ModelLong}\neq {\rm `-.-'}$,
                 ${\rm ModelLat}\neq {\rm `-.-'}$, and ${\rm Sigma}\neq {\rm `-.-'}$).\\
{\tt Q}     & 52    & I1   & Subjective quality, all directly connected to the ecliptic drawn by Scheiner get
                 $Q=1$. The rotated sunspot groups (cf.~Fig.~\ref{scheiner_035}) are probably slightly less accurate and get $Q=2$.
                 Positions derived from rotational matching may also obtain $Q=2$ or 3, if the
                 probability distributions fixing the position angle of the drawing were not
                 very sharp, or broad and asymmetric, respectively. Methods `H' and `A' always 
                 obtain $Q=3$, because of the assumptions made.
                 Spots for which no position could be derived, but have sizes, get $Q=4$.\\
{\tt SS}       & 54--55  & I2   & Size estimate in 13 classes running from 1 to 13. The classes are different from the ones used in \citet{arlt_ea2013} and \citet{pavai_ea2015} by the fact that we introduced a smaller size at the low end and named it ``1''. The classes are arbitrary anyway.\\
{\tt GROUP}    & 57--64  & C8   & Arbitrary group name; the order of numbers has no meaning\\
{\tt MEASURER} & 66--75  & C10  & Last name of person who obtained position\\
{\tt MOD\_L}   & 77--81  & F5.1 & Model longitude from rotational matching (only spots used for matching have this)\\
{\tt MOD\_B}   & 83--87  & F5.1 & Model latitude from rotational matching (only spots used for matching have this)\\
{\tt SIGMA}    & 89--93  & F5.3 & Total residual  of model positions compared with measurements of reference spots in rotational matching (only spots used for the matching have this). Holds for entire day.\\
{\tt DELTA}    & 95--98  & F4.1 & Heliocentric angle between the spot and the apparent disk centre in degrees (disk-centre distance); it is NaN if the spot position could not be determined.\\
{\tt UMB}      &100--103 & I4   & Umbral area in millionths of the solar hemisphere (MSH), corrected for foreshortening; it is NaN if spot position could not be derived.\\
{\tt A}        &105      & C1   & Flag (`!') marking areas which are highly uncertain since the spots appear to be drawn at too large sizes.\\
\hline
\end{tabular}
\tablefoot{The Format column uses the following designations: I denotes 
integer fields with the number behind being the number of characters; similarly, C is a character text field with the corresponding length, and, e.g. F5.1 is a floating point field of five characters length with one decimal.}
\end{table*}

\begin{figure}
\centering
\includegraphics[width=0.485\textwidth]{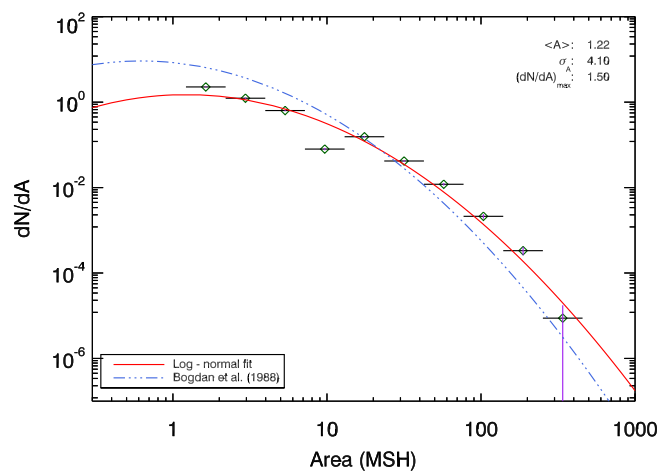}
\caption{Area distribution of 5555~spots drawn by Christoph Scheiner
in 1618--1631, all within a central-meridian distance of 
$|{\rm CMD}|\leq 50\degr$.\label{areadistribution}}
\end{figure}


\begin{figure*}
\centering
\includegraphics[width=0.98\textwidth]{butterfly.ps}
\caption{Butterfly diagram of the sunspot positions obtained from the
observations by Christoph Scheiner and his colleagues. The ordinate is
linear in $\sin \lambda$. The umbral areas of the individual spots are
used to weight the increments accumulating in each time and latitude bin. The 
approximate activity minimum is the time recently inferred by 
\cite{neuhaeuser2016}.\label{butterfly}}
\end{figure*}

The areas were taken from pixel counts in the individual
cursor shapes and converted into fractions of the solar
disk. They are then corrected for foreshortening and given
in millionths of the solar hemisphere (MSH). We follow the
procedure by \citet{bogdan_ea1988} to compute the area
distribution of the individual umbrae. The result is shown
in Fig.~\ref{areadistribution} and shows a fairly good
agreement with their distribution. We interpret the differences
from the curve obtained by \citet{bogdan_ea1988} as follows.
Firstly, large spots in particular may be exaggerated slightly in size, an effect 
seen in many other historic sunspot drawings. At the same time
several of the smallest spots were missed due to the observational
limits (chromatic telescope).
These two effects explain the slight overabundance of spots at
large areas and the slight under-abundance of spots between 
1.2~and 10~MSH. A correction factor of 0.8 may be suitable
for the areas, leading to a perfect match with the curve by 
\citet{bogdan_ea1988} at large areas. 
However, since this is speculation, we have not used any
correction factor for the areas in the resulting data file.
It would bring the smallest spots to even smaller areas
which were very likely not observable.

The smallest spot measurement delivered an area of only 1.2~MSH.
A circular spot of this area has an angular extent of $3''$ in
the center of the solar disk. It may be doubted that the
resolving capabilities of Scheiner's telescopes were as
good as that. While \citet{king1955} reported about Galilei's 
largest telescope having an aperture of 5.1~cm, but usually 
being stopped down to 2.6~cm. \citet{abetti1954} reports 
about a test of one of the first telescopes of Galileo and 
found a resolving power no better than $10''$. By the 1620s, 
the telescopes may have had improved, and Scheiner also used
a Keplerian setup providing better image quality.
We found only the following remark about
the necessary telescope for precise enough following of the
sunspots' motion in \citet[][p.~129-2]{scheiner1630}:
\begin{quote}
We obtain this to
an extraordinary degree, if the lens is not only a portion of a large
sphere, whose radius spans 20, 30, or
more Roman palms, but also if the same [lens] is sufficiently wide,
of at least one or two palms; so it will lend itself to do
something adequate, as long as it is made of material of good quality
and shaped without defects.
\end{quote}
The Roman palm measured 74~mm \citep{brockhaus1991} leading to 
impressive suggested lens diameters of 7--15~cm. We did not find an
exact aperture of the actual telescopes used, but conclude
that the resolution in the 1620s was significantly better than the one
of Galilei's early telescopes. Since the drawings -- especially the ones
of 1624--1631 do contain very small spots, we decided to
preserve the size information and leave possible recalibrations
of the areas to future applications of the sunspot data.

\section{Data format and butterfly diagram}\label{results}
We use the same table format as the one employed for
the spot positions and sizes derived from the observations 
by Schwabe \citep{arlt_ea2013}. The format is detailed in 
Table~\ref{format}. While the measurements deliver
central-meridian distances on the Sun, heliographic longitudes
are obtained using the solar disk center provided by the JPL
Horizons ephemeris generator
seen from the location in Rome (differences to the various
other geographical positions in Europe are extremely tiny).

Since Scheiner drew the sunspots -- at first glance -- to scale 
in the years of 1618--1631, we computed physical areas in
microhemispheres (MSH), corrected for the projection effect
towards the solar limb. The observations of 1611--1612 do not
show realistic areas. We chose cursor sizes of roughly half
the diameter of the spot in the image in order to compensate
for the exaggerated spots to some degree subjectively. Nevertheless,
these areas should be used with extreme care.

We define sunspot groups based on Scheiner's drawings
and assigned group numbers. The order of the group numbers
has no meaning; for technical reasons they are not ascending
monotonically. The groupings have
been more difficult for 1611, since the groups may have been
drawn too large in size. The remaining group definitions were
fairly straight-forward, since the drawings have a look similar
to modern white-light images (except for faculae). The
differences to the group definitions made already by Scheiner
are not too large. A total of 16 groups were split into two
groups, while 10~groups are the result of combining groups together.
Those numbers increased slightly as compared to 
\citet{pavai_ea2016} after another careful inspection.

A total of about 8152~spot positions were derived for
1611--1631. The exact number may still evolve upon
further investigations, since the distinction between
spots and faculae is not unambiguous (also drawn by
dark ink). The resulting butterfly diagram is shown in
Fig.~\ref{butterfly}. A considerable part of a cycle
is covered only in the years 1624--1631 where the
migration of spot emergences appears to be equatorward.
Following \citet{zolotova_ponyavin2015}, we name this
cycle $-12$. The general migration of sunspot emergence
latitude towards the equator is clearly seen. The
onset of cycle $-12$ took place first in the southern 
hemisphere, while no spot was found in the northern 
hemisphere. During the second half of cycle~$-12$,
the average latitudes of the northern wing are farther away
from the equator than in the southern one. The positions
are compatible with a time of minimum of fall 1620 as
suggested by \citet{neuhaeuser2016}. \citet{spoerer1889}
gives 1619, while \citet{hoyt_schatten1998} obtained
very low group sunspot numbers for both 1617 and 1618
which are mainly due to an overabundance of zero-detections
inferred from generic statements of Simon Marius and
Andrea Argoli (incorrectly reported as seen by Riccioli
in \citeauthor{hoyt_schatten1998} \citeyear{hoyt_schatten1998}).

Cycle~$-13$ is more poorly covered. The 1612 observations
show the presence of the two butterfly wings nicely, while
the 1611 positions are too inaccurate to exhibit the 
hemispheric division. A slight dominance of the northern
hemisphere may be detectable in both 1611 and 1612.

The spot positions and areas are publicly 
available at the astronomical data center CDS.
Since none of the observational sources before the Maunder
minimum covers a sufficient period in time to deliver useful
information about a full cycle -- the latitudinal distribution
of spots in the first place --, a unified database of the
various data sets will be very advantageous.

\section{Sunspot group tilt angles\label{tiltangles}}
The drawings by Scheiner and his colleagues were then 
manually inspected for potential bipolar sunspot groups.
We restricted the analysis to the very realistic
drawings of 1618--1631. The relevant groups were
flagged in the positional database and tilt angles
were computed according to the method described by
\citet{pavai_ea2015}. The data format of the tilt angle
data file is exactly the same as in that paper. The
total number of tilt angles obtained is 697.

\begin{figure}
\centering
\includegraphics[width=0.485\textwidth]{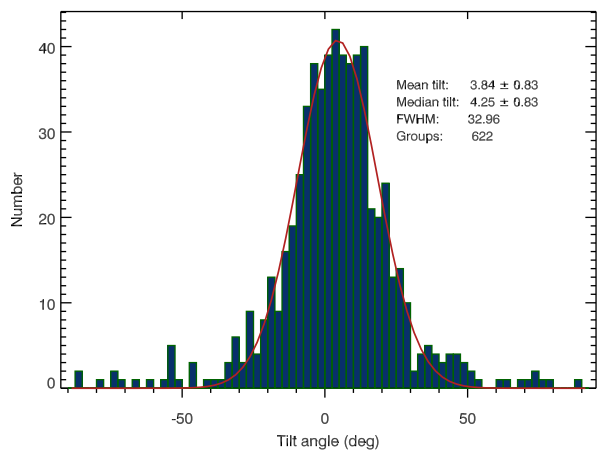}
\caption{Tilt angle distribution for the observations by Scheiner
in 1618--1631. The solid line is a Gaussian fit to the distribution
and delivers the full width at half-maximum (FWHM). (After
\citet{pavai_ea2016})
\label{tiltdist}}
\end{figure}

The distribution of the 622~tilt angles within central meridian
distances of $|{\rm CMD}|<60\degr$ is
shown in Fig.~\ref{tiltdist}. The width of the distribution
was derived from a Gaussian fit to the histogram. The
average tilt angle of $3\fdg84\pm0\fdg83$ is slightly
lower than values of the 20th century. Values by \citet{wang_ea2015}
for cycles~16--23 range from $4\fdg6$ to $6\fdg6$, if a
minimum polarity separation of $2\fdg5$ is used to separate
unipolar groups with more than one spot from true bipolar 
groups. Because of the relatively large error margin of
the average for Scheiner, we have to conclude that the
value is not significantly lower than in the 20th century
(see \citet{pavai_ea2016} for more average tilt angles in
the course of four centuries).
Most of the data are from
1625--1627 which is roughly the year of solar maximum 
and two years of the descending activity.

\section{Summary}\label{summary}
The solar disk drawings with sunspots made by Christoph
Scheiner and colleagues in 1611--1631 were digitized and
measured. The three sources for the drawings are \citet{scheiner1630}, \citet{scheiner1651},
and \citet{reeves2010}. A total of 8167~spot areas
were obtained of which 8152~are accompanied by heliographic 
positions. All measurements are provided in a database file. The accuracy 
of both positions and areas are poor for 1611. The
positional accuracy improved in the 1612 observations but
the spot areas are still highly exaggerated.
High quality drawings of 1618--1631 delivered a 
positional accuracy of about $1\degr$--$2\degr$ in heliographic 
coordinates in the solar disk centre, thanks to the large
scale of the drawings. The database does not contain spotless
days. We refer to the detailed tables by \citet{hoyt_schatten1998}
for estimates of the spotless days which go beyond what 
\citet{scheiner1630} reported.

Sunspot numbers may also be incomplete as indicated by
two groups in \citet[][Plate I, belonging to p.~7]{scheiner1651} 
seen on 1625 May~23--29 and 1626 Jun~30--Jul~12, respectively, 
which were not shown in the images of \citet{scheiner1630}. 
Since this is the only image with overlap between the two 
books, it is not possible to estimate the general completeness.

The positional data support the migration of sunspot emergences
towards the equator through cycles~$-13$ and $-12$. Apart from
the very inaccurate 1611 data, there were two groups in
1629 which straddle the equator. On some days, just a few 
small spots are on the other hemisphere, but on two days (one
day for each group), the average polarities sit in opposite 
hemispheres. Near-equator groups may be interesting for the 
progress of the activity cycle as recently suggested by 
\citet{cameron_ea2013}. For the accurate period of 1618--1631, we find
18~bipolar groups having a group center latitudes of
$|\lambda|\leq5\degr$.

We computed 697~sunspot group tilt angles from a manually selected
set of supposedly bipolar group instances (i.e. the same group may 
be used on more than one day) of 1618--1631 and provide them in a 
separate data file. The average tilt angle for these 
observations is $3\fdg84\pm0\fdg83$ and is not significantly
different from 20th-century values, albeit on the low side.
There were 1341~group instances not selected as being bipolar in 1618--1631.

The data will be made available at the astronomical data center
CDS.\footnote{Copies of the positional data file and the tilt 
angle data file will be available at {\tt http://www.aip.de/Members/rarlt/sunspots}.}


\begin{acknowledgements}
We are grateful to Regina v. Berlepsch for digitizing the Rosa Ursina 
by Christoph Scheiner in the library of Leibniz Institute for Astrophysics 
Potsdam, to Nadya Zolotova from Sankt Petersburg for drawing our attention
to \citet{scheiner1651}, to Ralph Neuh\"auser from Jena University for many helpful comments, 
and to Daniela Luge from Jena University for translating some Latin phrases. 
VSP thanks the Deutsche Forschungsgemeinschaft for the support
in grant Ar\,355/10-1. 
\end{acknowledgements}


\bibliographystyle{aa}
\bibliography{scheiner}

\begin{thebibliography}{36}
\expandafter\ifx\csname natexlab\endcsname\relax\def\natexlab#1{#1}\fi

\bibitem[{{Abarbanell} \& {W{\"o}hl}(1981)}]{abarbanell_woehl1981}
{Abarbanell}, C. \& {W{\"o}hl}, H. 1981, \solphys, 70, 197

\bibitem[{{Abetti}(1954)}]{abetti1954}
{Abetti}, G. 1954, {The History of Astronomy} (London: Sidgwick and Jackson)

\bibitem[{{Apelles}(1612)}]{scheiner1612}
{Apelles}, C.~S. 1612, {De maculis solarib[us] et stellis circa Iovem
  errantibus, accuratior disquisitio} (Augsburg: Ad insigne pinus)

\bibitem[{{Arlt} {et~al.}(2013){Arlt}, {Leussu}, {Giese}, {Mursula}, \&
  {Usoskin}}]{arlt_ea2013}
{Arlt}, R., {Leussu}, R., {Giese}, N., {Mursula}, K., \& {Usoskin}, I.~G. 2013,
  \mnras, 433, 3165

\bibitem[{{Balthasar} {et~al.}(1986){Balthasar}, {V\'azquez}, \&
  {W\"ohl}}]{balthasar_ea1986}
{Balthasar}, H., {V\'azquez}, M., \& {W\"ohl}, H. 1986, \aap, 155, 87

\bibitem[{{Birkenmajer}(1967)}]{birkenmajer1967}
{Birkenmajer}, A. 1967, Vistas in Astronomy, 9, 11

\bibitem[{{Bogdan} {et~al.}(1988){Bogdan}, {Gilman}, {Lerche}, \&
  {Howard}}]{bogdan_ea1988}
{Bogdan}, T.~J., {Gilman}, P.~A., {Lerche}, I., \& {Howard}, R. 1988, \apjl,
  327, 451

\bibitem[{{Braunm\"uhl}(1891)}]{braunmuehl}
{Braunm\"uhl}, A. 1891, {Christoph Scheiner als Mathematiker, Physiker und
  Astronom} (Bamberg: Buchnersche Verlagsbuchhandlung)

\bibitem[{{Brockhaus}(1991)}]{brockhaus1991}
{Brockhaus}. 1991, {Brockhaus-Enzyklop\"adie, Vol.~16, Nos--Per} (Mannheim:
  F.A.~Brockhaus)

\bibitem[{{Brockhaus}(1992)}]{brockhaus1992}
{Brockhaus}. 1992, {Brockhaus-Enzyklop\"adie, Vol.~19, Rut--Sch} (Mannheim:
  F.A.~Brockhaus)

\bibitem[{{Cameron} {et~al.}(2013){Cameron}, {Dasi-Espuig}, {Jiang}, {I{\c
  s}{\i}k}, {Schmitt}, \& {Sch{\"u}ssler}}]{cameron_ea2013}
{Cameron}, R.~H., {Dasi-Espuig}, M., {Jiang}, J., {et~al.} 2013, \aap, 557,
  A141

\bibitem[{{Casas} {et~al.}(2006){Casas}, {Vaquero}, \&
  {Vazquez}}]{casas_ea2006}
{Casas}, R., {Vaquero}, J.~M., \& {Vazquez}, M. 2006, \solphys, 234, 379

\bibitem[{{Eddy} {et~al.}(1976){Eddy}, {Gilman}, \& {Trotter}}]{eddy_ea1976}
{Eddy}, J.~A., {Gilman}, P.~A., \& {Trotter}, D.~E. 1976, \solphys, 46, 3

\bibitem[{{Eddy} {et~al.}(1977){Eddy}, {Gilman}, \& {Trotter}}]{eddy_ea1977}
{Eddy}, J.~A., {Gilman}, P.~A., \& {Trotter}, D.~E. 1977, Science, 198, 824

\bibitem[{{Harriot}(1613)}]{harriot1613}
{Harriot}, T. 1613, {Spots on the Sun} (Petworth House, HMC 241 VIII,
  http://echo.mpiwg-berlin.mpg.de/MPIWG:FAYG83FB)

\bibitem[{{Herr}(1978)}]{herr1978}
{Herr}, R.~B. 1978, Science, 202, 1079

\bibitem[{{Hockey} {et~al.}(2007){Hockey}, {Trimble}, {Williams}, {Bracher},
  {Jarrell}, {March{\'e}}, {Ragep}, {Palmeri}, \& {Bolt}}]{hockey_ea2007}
{Hockey}, T., {Trimble}, V., {Williams}, T.~R., {et~al.} 2007, {The
  Biographical Encyclopedia of Astronomers} (New York: Springer)

\bibitem[{{Hoyt} \& {Schatten}(1998)}]{hoyt_schatten1998}
{Hoyt}, D.~V. \& {Schatten}, K.~H. 1998, \solphys, 181, 491

\bibitem[{{King}(1955)}]{king1955}
{King}, H.~C. 1955, {The history of the telescope} (New York: Dover and London:
  Griffin)

\bibitem[{{Landsman}(1993)}]{landsman1993}
{Landsman}, W.~B. 1993, in Astronomical Society of the Pacific Conference
  Series, Vol.~52, Astronomical Data Analysis Software and Systems II, ed.
  R.~J. {Hanisch}, R.~J.~V. {Brissenden}, \& J.~{Barnes}, 246

\bibitem[{{Malapertius}(1620)}]{malapert1620}
{Malapertius}, C. 1620, {Oratio habita Duaci dum lectionem mathematicam
  auspicaretur} (Douai: Balthasar Beller)

\bibitem[{{Malapertius}(1633)}]{malapert1633}
{Malapertius}, C. 1633, {Austriaca sidera heliocyclia astronomicis hypothesibus
  illigata} (Douai: Balthasar Beller)

\bibitem[{{Neuh{\"a}user} \& {Neuh{\"a}user}(2016)}]{neuhaeuser2016}
{Neuh{\"a}user}, R. \& {Neuh{\"a}user}, D.~L. 2016, Astronomische Nachrichten,
  337, 581

\bibitem[{{Reeves} \& {Van Helden}(2010)}]{reeves2010}
{Reeves}, E. \& {Van Helden}, A. 2010, {On sunspots. Galileo Galilei and
  Christoph Scheiner} (University of Chicago Press)

\bibitem[{{Rek}(2010)}]{rek2010}
{Rek}, R. 2010, \solphys, 261, 337

\bibitem[{{Scheiner}(1630)}]{scheiner1630}
{Scheiner}, C. 1630, {Rosa Ursina sive Sol} (Bracciano: Andreas Phaeus)

\bibitem[{{Scheiner}(1651)}]{scheiner1651}
{Scheiner}, C. 1651, {Prodromus pro sole mobili et terra stabili} (Nysa,
  Silesia: Collegium Nissense Societatis Iesu)

\bibitem[{{Senthamizh Pavai} {et~al.}(2015){Senthamizh Pavai}, {Arlt},
  {Dasi-Espuig}, {Krivova}, \& {Solanki}}]{pavai_ea2015}
{Senthamizh Pavai}, V., {Arlt}, R., {Dasi-Espuig}, M., {Krivova}, N.~A., \&
  {Solanki}, S.~K. 2015, \aap, 584, A73

\bibitem[{{Senthamizh Pavai} {et~al.}(2016){Senthamizh Pavai}, {Arlt},
  {Diercke}, {Denker}, \& {Vaquero}}]{pavai_ea2016}
{Senthamizh Pavai}, V., {Arlt}, R., {Diercke}, A., {Denker}, C., \& {Vaquero},
  J.~M. 2016, Advances in Space Research, in press

\bibitem[{{Smogulecz} \& {Sch\"onberger}(1626)}]{smogulecz_schoenberger1626}
{Smogulecz}, J. \& {Sch\"onberger}, G. 1626, {Sol illustratus ac propugnatus}
  (Freiburg im Breisgau: T. Meyer)

\bibitem[{{Sp\"orer}(1889)}]{spoerer1889}
{Sp\"orer}, G. 1889, {Ueber die Periodicit\"at der Sonnenflecken seit dem Jahre
  1618} (Leipzig: Wilh. Engelmann)

\bibitem[{{Vaquero} \& {Trigo}(2015)}]{vaquero_trigo2015}
{Vaquero}, J.~M. \& {Trigo}, R.~M. 2015, \na, 34, 120

\bibitem[{{Wang} {et~al.}(2015){Wang}, {Colaninno}, {Baranyi}, \&
  {Li}}]{wang_ea2015}
{Wang}, Y.-M., {Colaninno}, R.~C., {Baranyi}, T., \& {Li}, J. 2015, \apj, 798,
  50

\bibitem[{{Yallop} {et~al.}(1982){Yallop}, {Hohenkerk}, {Murdin}, \&
  {Clark}}]{yallop_ea1982}
{Yallop}, B.~D., {Hohenkerk}, C., {Murdin}, L., \& {Clark}, D.~H. 1982, \qjras,
  23, 213

\bibitem[{{Zinner}(1957)}]{bio}
{Zinner}, E. 1957, {in {Graf zu Stolberg-Wernigerode}, O. et al. (eds.): Neue
  deutsche Biographie. Dritter Band} (Berlin: Duncker \& Humblot)

\bibitem[{{Zolotova} \& {Ponyavin}(2015)}]{zolotova_ponyavin2015}
{Zolotova}, N.~V. \& {Ponyavin}, D.~I. 2015, \apj, 800, 42

\end{thebibliography}

\end{document}